\documentclass[doublecol,aps,epl,preprintnumbers,nofootinbib,floatfix]{epl2}
\usepackage{graphicx,color,amsmath}
\usepackage{hyperref}
\usepackage{amsfonts}
\usepackage{amsmath}
\usepackage{listings}
\usepackage{hyperref}
\usepackage[T1]{fontenc}
\usepackage[english]{babel}

\begin{document}

\title{Electrostatic interactions in the presence of surface charge
  regulation: exact results} \author{A. C. Maggs \inst{1} \and
  R. Podgornik \inst{2}} \shortauthor{Maggs and Podgornik}
\shorttitle{Surface charge regulation}

\institute{\inst{1} Physico-chimie th\'eorique--Gulliver, ESPCI-CNRS,
  10 rue Vauquelin 75005 Paris, France \\ \inst{2} Department of
  Theoretical Physics, J. Stefan Institute and Department of Physics,
  Faculty of Mathematics and Physics, University of Ljubljana -
  SI-1000 Ljubljana, Slovenia}

\pacs{82.70.Dd}{Colloids} \pacs{87.10.+e}{General, theoretical, and
  mathematical biophysics} \date{}

\abstract {We study the problem of charge regulation and its effects
  on electrostatic interactions between dissociable charge groups
  immersed in a univalent electrolyte, within a family of one
  dimensional exactly solvable models. We consider the case of both
  charge regulated plates, but also the interaction of pairs of finite
  size dielectric "particles". Using the transfer matrix formalism we
  are able to determine the disjoining pressure as well as the
  correlations between the charge and the dipole moments of the
  objects as a function of their separation and electrolyte
  concentration.}

\maketitle

\section{Introduction}
The interaction of charged objects immersed in an electrolyte gives
rise to a rich phenomenology which has been explored by a number of
different methods \cite{perspective}. In this paper we study
one-dimensional models using a field theory formalism \cite{funint,
  netzorland} which was introduced by Edwards and Lenard
\cite{edwardsLenard}, and adapted to surface properties and
surface-surface interactions in \cite{dean,dean2} and
\cite{rudisurf,book,vincent}.

We are particularly interested in potential applications for the
interactions between proteins \cite{biochargereg}, where it has been
realised for a long time that fluctuations in surface amino acid
charge dissociation state can give rise to a monopole-monopole
fluctuation interaction {as described by Kirkwood and Shumaker}
\cite{kirkwood1, kirkwood2}. Such a long-range fluctuation interaction
is possible only for surfaces that exhibit charge regulation, that is,
they do not have a fixed surface charge. {When present these
  interactions decay, in three dimensions, as $1/r^2$ between point
  particles and vary consequently as $\log{(D)}$ between planar
  surfaces \cite{rudisurf}.}  In the present paper we study effects of
this type {\sl exactly} in one dimension by using the surface
dissociation model introduced by Ninham and Parsegian \cite{ninham} on
the mean-field level. The model was developed further in different
contexts \cite{Chan, Szleifer, Boon, Borkovec, netzchargereg} and
recently reformulated in terms of the free energy functions for
dissociable surface charges \cite{rudisurf}.

We base our analysis on the functional integral representation of the
partition function generalized by the inclusion of a surface effective
free energy that describes the charge regulation. The model is solved
exactly and the results provide insights into the correlations between
the charge states of interacting surfaces. We go a step further and
consider the interaction of two charged "particles" with a dielectric
core and dissociable surface charge groups.

\section{One dimensional electrolytes}
We consider a one dimensional system of positive and negative charges
bounded by two charged interfaces located at $x = 0, L$.  The charges
interact with the unidimensional Coulomb interaction:
\begin{equation}
  H=-\frac{1}{4}\sum_{ij} | x_i -x_j | q_i q_j = -\frac{1}{2}\sum_{ij}
  G_c(x_i, x_j) q_i q_j,
\end{equation}
in units where the dielectric constant is unity.  $x_i$ is the
position of the particle $i$ carying charge $q_i$ and the Coulomb
kernel $G_c(x, x')$ satisfies
\begin{equation}
  \frac{\partial^2 G_c(x, x')}{\partial x^2} = -\delta(x-x'). 
\end{equation}
We can then use field theory mappings \cite{funint,netzorland}, or
analogies with a one-dimensional Brownian particle \cite{dean}, to
show that the free energy functional for an interacting symmetric
one-one electrolyte is
\begin{equation}
  S[\phi] =\int dx \bigg(   {\textstyle\frac{1}{2}} \beta  \left( \frac{\partial
      \phi} {\partial x} \right)^2 {-} 2 \lambda \cos(\beta e \phi) + i
  \rho_0 \phi\bigg)
  \label{bcefghjs}
\end{equation}
where $\beta$ is the inverse thermal energy, {$\lambda$ is the
  absolute activity of the electrolyte bath}, $e$ is the elementary
charge, $\rho_0$ is an external fixed charge density and $\phi$ is a
fluctuating potential which is integrated over to calculate the
partition function (up to an irrelevant multiplicative constant)
\begin{equation}
  Z= 
  \int {\mathcal D}[\phi]~e^{-S[\phi]}, \label{equ1}
\end{equation}
so $S[\phi]$ can be viewed appropriately as a field action. The
prefactor of the functional integral is irrelevant for the specific
context of this work.

\begin{figure}[ht]
  \includegraphics[scale=.27] {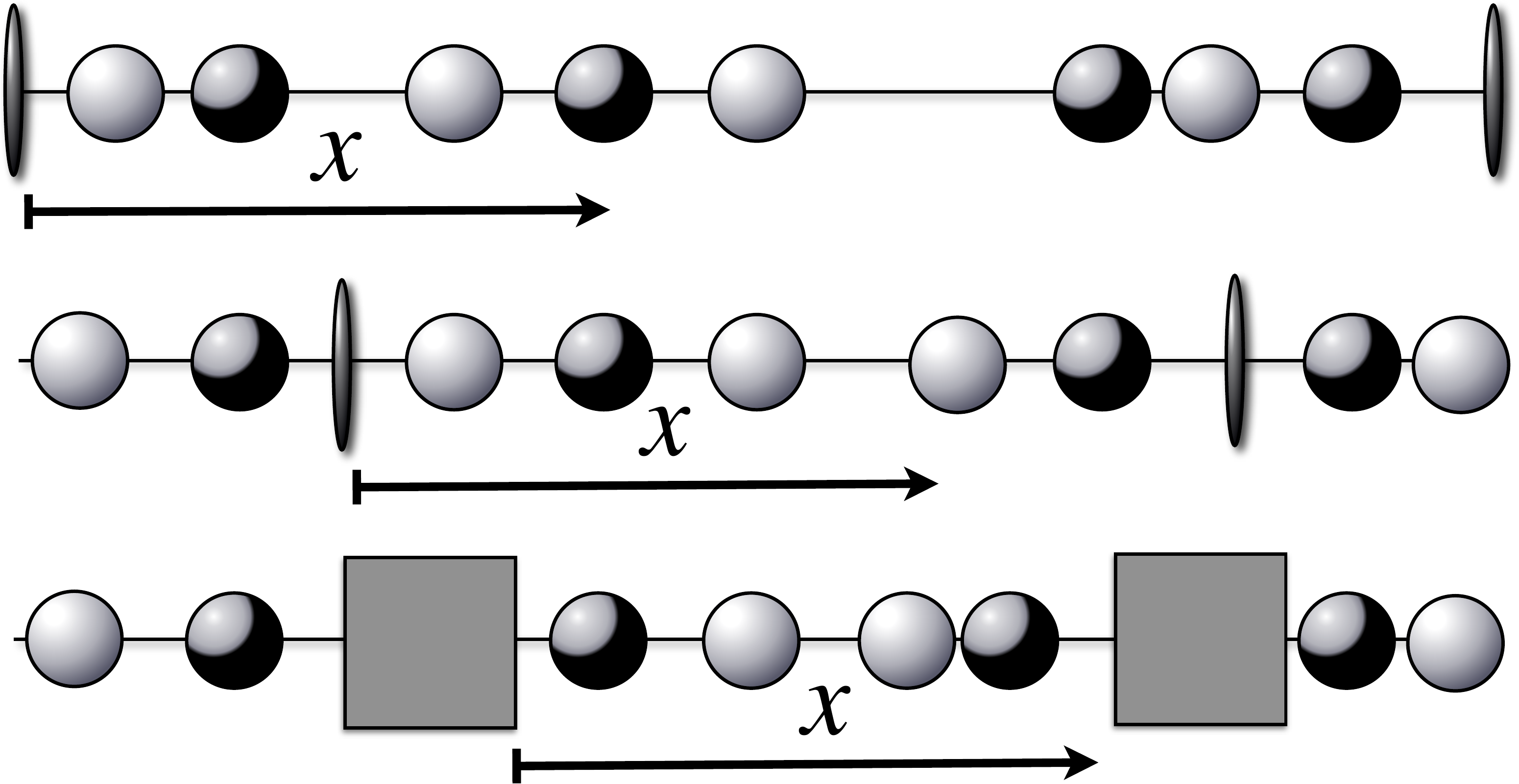}
  \caption{Schematic representation of the three considered
    systems. Top: two metallic interacting ionizable surfaces.
    Middle: two ionizable surfaces in an infinite electrolyte
    bath. Bottom: two dielectric "particles" with ionizable surfaces
    in an infinite electrolyte bath.}
  \label{interactingplate}
\end{figure}

By rescaling the energy and the length scales in the problem we can
work with an a-dimensional form of the interaction. We chose to scale
the variables such that the bulk field action takes the form
\begin{equation}
  S[\psi] = {\textstyle\frac{1}{2}}\int _0^L dt \left ( \frac{d \psi }{dt} \right)^2
  - 2 \bar \lambda \int_0^L dt \cos(\psi)
\end{equation}
From now on we will use this form, measuring distances in units of the
one-dimensional Bjerrum length $\ell_B=1/e^2\beta$ and the fluctuating
{field action} in thermal units. Note there is an inversion in the
physics of one-dimensional electrolytes, compared to three
dimensions. Particles interact weakly at small separations and
strongly at the largest distances; the potential strength increasing
with $|x_i - x_j|$. This implies that dense electrolytes, with large
$\bar \lambda $ are described by the simpler weakly interacting
theory. For small $\bar \lambda $ particles can form interacting,
bound Bjerrum pairs \cite{edwardsLenard}.

The remarkable feature of the one dimensional partition function that
we will exploit \cite{edwardsLenard} is that the non-linear weighting
eq.~(\ref{equ1}) maps onto another {\em linear}\/ problem. This
mapping is essentially the same as the mapping from the path-integral
to the Schr\" odinger formulations of quantum mechanics. We will thus
be concerned with linear partial differential equations of the form
\begin{equation}
  \frac{\partial G_{\bar\lambda}(\psi, \psi'; t)}{\partial t}= {\textstyle\frac{1}{2}}
  \frac{\partial^2 G_{\bar\lambda}(\psi, \psi'; t)}{\partial \psi^2} + 2 \bar \lambda
  \cos(\psi) G_{\bar\lambda}(\psi, \psi'; t)
\end{equation}
where the time like variable $t$ is the position in the
one-dimensional electrolyte measured in terms of the Bjerrum length
and $G_{\bar\lambda}(\psi, \psi'; t)$ is a Green's function.

\section{Surface free energy}

If external charges on the bounding surfaces $x=0, L$ are considered
to be equal and fixed at $N e$, then they contribute
\begin{equation}
  i \rho_0 \phi \longrightarrow i Ne \phi(0) \delta(x) + i Ne \phi(L) \delta( x - L)
\end{equation}
to the field action. However, we rather consider the case where each
surface can be in a state of charge which varies from $-N e$ to $(n_s
-N) e$, corresponding to a surface charge group dissociation
equilibrium. The chemical potential for dissociation is assumed to be
independent of the number of previously bound ions, corresponding to
the {\sl charge regulation}\/ paradigm \cite{rudisurf}.  On mapping to
the field theory form we find that the surface free energy is given by
the surface lattice gas expression:
\begin{equation}
  e^{-\beta f(\psi)} = e^{-iN \psi +n_s \ln(1+\lambda_S e^{i \psi})}.
\end{equation}
where $\log{\lambda_S}$ is proportional to the free energy cost of
charge dissociation.  It will be useful to expand the logarithmic term
into a {\em finite sum}\/ to find
\begin{align}
  e^{-\beta f(\psi)} =& e^{-iN \psi} (1+ \lambda_s e^{i\psi})^{n_s}\\
  =& \sum_{k=0}^{n_s} {n_s \choose k }\lambda_s^{n_s -k }
  e^{i(n_s-k-N))}.
\end{align}
In our numerical applications we make the choice $n_s=2N$ so that the
surface can have a charge state which varies between $-N e$ to $+N e$,
but other choices are possible.

The total partition function of two surfaces, separated by $L$ in
Bjerrum length units, interacting through a one-one electrolyte is
then \cite{dean}
\begin{equation}
  \Xi(L) = 
  \int d\phi_0 \int d\phi_L \; e^{-\beta f(0)} G_{\bar\lambda}(\phi_0,
  \phi_L; L) e^{-\beta f(L)}.
  \label{partfun}
\end{equation}
This implies our principal result relating the partition function to a
finite sum over Green functions:
\begin{align}
  \Xi(L) = &\sum_{k,k'=0}^{n_s} {n_s \choose k}{n_s \choose k'}
  \lambda_s^{2n_s-k-k'} \times \nonumber \\ &\int d \psi \; d\psi'\;
  e^{iM\psi} G_{\bar\lambda}(\psi, \psi'; L) e^{iM' \psi'}
\end{align}
with $M=n_s-N-k$, $M'=n_s-N-k'$. Let us consider the Fourier
transformed Green function
\begin{align}
  Z_{M,M'}(L) = &\int d \psi \; d\psi'\; e^{iM\psi}
  G_{\bar\lambda}(\psi, \psi'; L) e^{iM' \psi'} \nonumber \\
  =&\int_0^{2 \pi} d \psi \; e^{iM \psi} K(\psi,L).
\end{align}
$K$ can be expanded as a Fourier series
\begin{equation}
  K(\psi,t) = \sum_{n=-\infty}^\infty  b(n,t) e^{i n \psi}.
\end{equation}

\begin{equation}
  \dot b_{nm}(t) = - {\textstyle\frac12}n^2 b_{nm}(t) +
  \bar\lambda ( b_{n+1,m}(t) + b_{n-1,m}(t)). \label{eq:b}
\end{equation}

\begin{figure*}[t!]\begin{center}
    \begin{minipage}[b]{0.495\textwidth}\begin{center}
        \includegraphics[width=\textwidth]{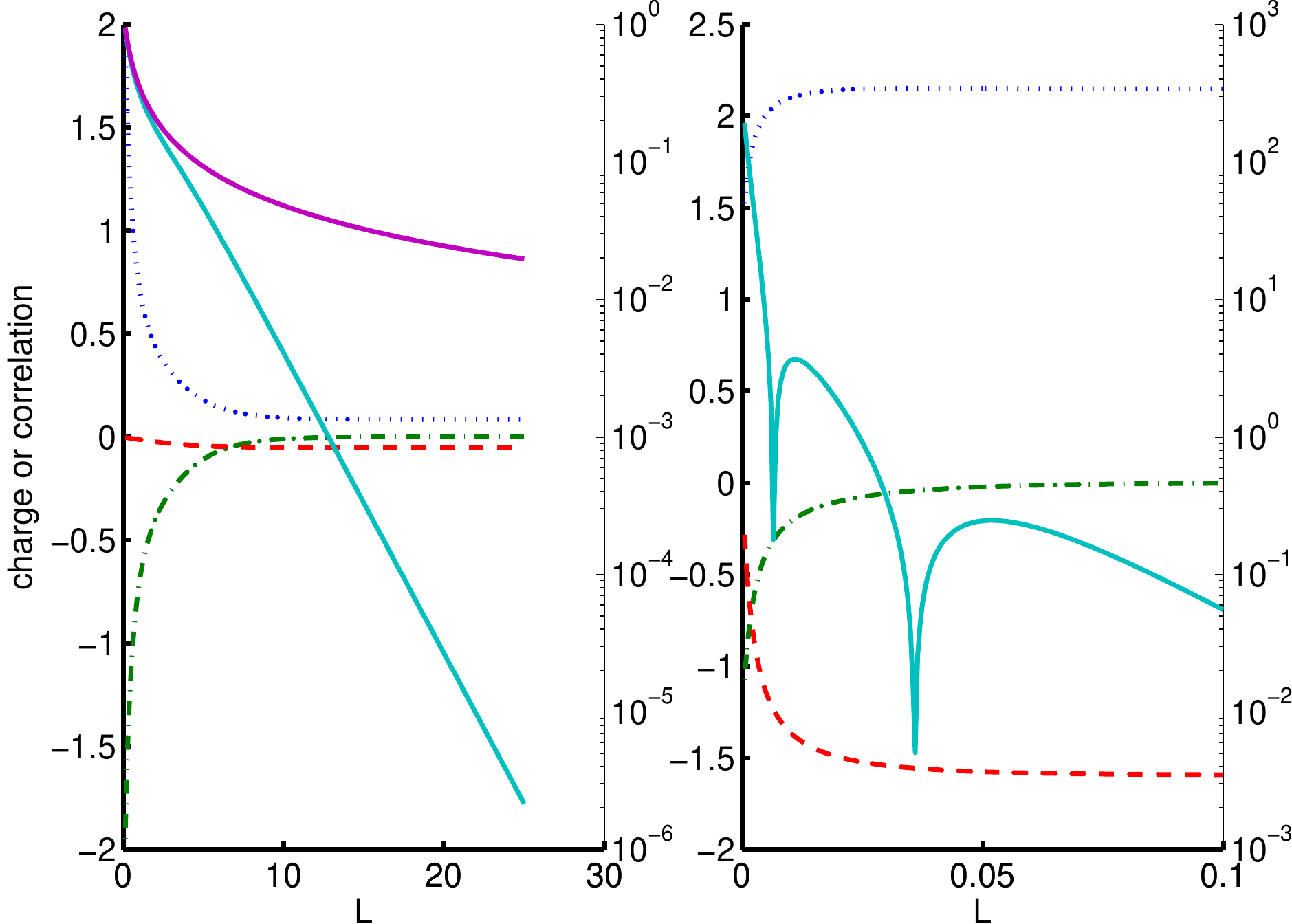}\\
        (a)~~~~~~~~~~~~~~~~~(b)
      \end{center}\end{minipage}\hskip0.1cm~~
    \begin{minipage}[b]{0.48\textwidth}\begin{center}
        \includegraphics[width=\textwidth]{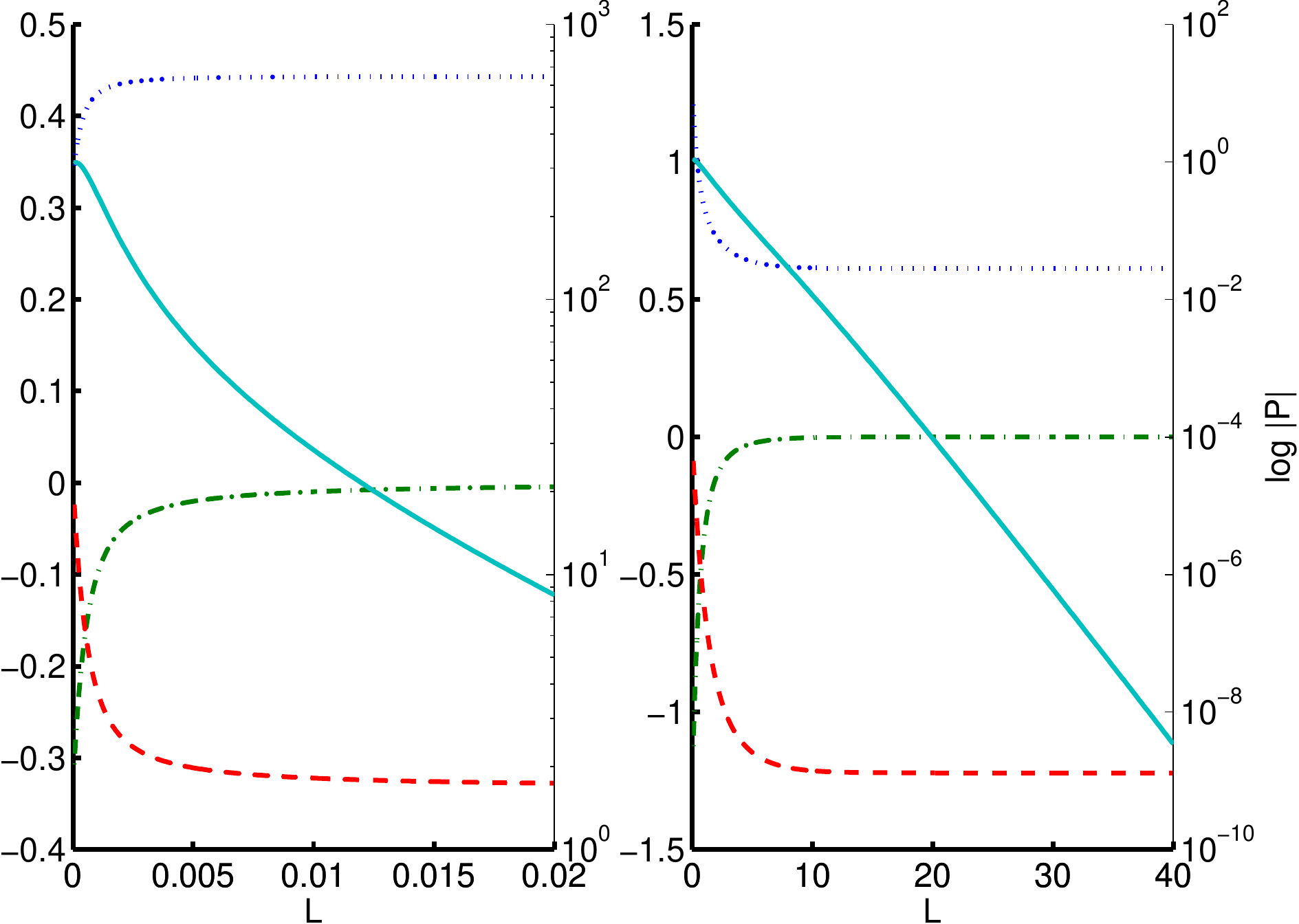}
        (c)~~~~~~~~~~~~~~~~~(d)
      \end{center}\end{minipage}\hskip0.1cm
    \caption{({\bf a}): Demonstration of the Kirkwood-Shumaker
      effect for two plates. {Turquoise}, solid line -- Log of the
      disjoining pressure as a function of dimensionless separation
      (attractive interactions). Red dashed -- average charge on a
      plate, blue dotted -- mean square fluctuation of charge on a
      single plate, green dashed-dotted -- charge correlation between
      two plates. $\bar \lambda=0.02$ and $\lambda_S=2$ with $N=10$;
      {top mauve line analytic expression eq.~(\ref{eq:P}) valid
        for small separations}.  ({\bf b}): 
      Parameters for two plates demonstrating
      intermediate range attraction, but repulsion for small and large
      separations.
      $\lambda_S=2$, $\bar \lambda=10^3$, $N=5$. This set of parameters leads to a
      minimum in the free energy at a finite separation between
      plates. 
      ({\bf c}): $N=1$, $\lambda_S=2$, $\bar \lambda=10^3$. $N=1$,
      plates are attracting.  ({\bf d}): $\bar \lambda=0.1$ and
      $\lambda_S=20$ with $N=5, n_s=10$, interaction is always
      repulsive. }
    \label{fig:1}
  \end{center}\end{figure*}

The initial condition is that \(b_{n,m}(0) = \delta_{n,-m}\).  This is
then a simple matrix equation $\dot b = \Gamma b $ with solution $b=
e^{\Gamma t} \tilde I$, where $\tilde I$ is a matrix that has unit
elements on the {\it skew diagonal}, corresponding to complex
conjugation.  The matrix \(\Gamma\) has entries \( -n^2/2\) on the
diagonal and entries \(\bar\lambda \) on the first step from the
diagonal. The thermodynamics of the system is then deduced from
\begin{equation}
  \Xi(L) = \sum_{k,k'=0}^{n_s} {n_s \choose k}{n_s \choose k'}
  \lambda_s^{2n_s-k-k'} b_{n_s-N-k , n_s-N-k'}(L),
\end{equation}
where $L$ is the extension of the system in the units of Bjerrum
length. The disjoining pressure $p$, the average charge and its mean
square fluctuations on the bounding surfaces, as well as the
cross-correlation of the charge between the two surfaces can all be
obtained straightforwardly from the above expression. {The
  disjoining pressure {in thermal units} follows from the standard
  expression in the form 
  \begin{equation}
    P = \frac{\partial}{\partial L} \log{\Xi(L)}. 
  \end{equation}
  The mean surface charge $\big\langle M \big\rangle $ and the
  mean-square surface charge fluctuation $\big\langle M^2 \big \rangle
  $ are identical at both bounding surfaces,
  with 
  $M = n_s-N-k$ at the boundary $x=0$, or equivalently at the boundary
  $x=L$ with $M \longrightarrow M' = n_s-N-k'$}, with the
average defined as
\begin{equation}
  \big\langle \cdots \big\rangle = \frac{1}{\Xi} \sum_{k,k'=0}^{n_s} {n_s \choose k}{n_s \choose k'}
  \lambda_s^{2n_s-k-k'} (\cdots)~b_{M, M'}(L).
\end{equation}

The cross-correlation function between charges on both surfaces, $\big
\langle \left( M - \big\langle M\big\rangle \right) \left( M' - \big<
  M'\big>\right)\big\rangle $, quantifies the correlation between the
instantaneous charge at boundary $x=0$ and boundary $x=L$ that depends on other
parameters and the size of the system.  {The disjoining pressure is
calculated within our code by the use of the relation 
\begin{equation}
P=\langle \Gamma \rangle.
\end{equation}
}

\section{Interactions without electrolyte}
We first consider the case of two metallic plates. We impose strict
electroneutrality on the system of plates so that the electric field
is identically zero outside of the considered region \cite{dean} and
start with the choice $\lambda_s=1$, {where the effects of monopole
  fluctuations can be expected to be strongest}. The case involving
$N=1$ is particularly simple. There are three surface charge states
possible: $(0,0)$, $(e,-e)$, $(-e,e)$. The partition function is then
\begin{math}
  Z = 1 + 2 e^{- \beta e^2 L/2 }
\end{math}
where we have used the energy $LE^2/2$ for the electric field with
$E=q$ as the boundary condition.  We find that the
pressure 
\begin{equation}
  P= \frac{e^2}{(2+ e^{\beta e^2L/2})}
\end{equation}
looks rather like a Fermi function, having finite values at small
separations and exponential decaying for separations beyond the
Bjerrum length. Charged states become exponentially rare at large
separations since their energy increases with separation. {This is a
  big difference with the three dimensional case where object can
  remain charged at large separations.}
  
There is also an interesting and simple result for a large number of
active sites ($N \gg 1$) in the limit $L<\ell_B$. We start with a
simple argument neglecting discrete charges. The energy of a system
with charge $q$ on a single plate is $E = q^2 L/2 $.  Thus
\begin{math}
  Z= \int dq e^{-\beta q^2 L/2} \sim \frac{1} {\sqrt{L} }
\end{math}
This gives the simple expression for the pressure
\begin{equation}
  P \sim  \frac{1}{2\beta } L^{-1}
\end{equation} 
While valid in the limit of very large $N$ it gives a rather poor fit
for moderate $N$. Much better results are obtained with the following
modified argument which includes the entropy of the charge
fluctuations.  The propagation of modes within the electrolyte is
given by eq.~(\ref{eq:b}):
\begin{math}
  b(t) = e^{-n^2 t/2} \tilde I.
\end{math}
The surfaces are then described by a Gaussian approximation to the
binomial coefficients for a near neutral surface with $n$ the net
number of charges:
\begin{math}
  {2N \choose{N+n}} \sim e^{-n^2/N} .
\end{math}
Then
\begin{equation}
  Z = \int dn\;  \left (e^{- n^2/N} \right )^2 e^{-n^2 L/2 \ell_B} \sim
  \frac{1}{\sqrt{L/\ell_B + 4 /N} }, \label{eq:P}
\end{equation}
giving:
\begin{equation}
  P = {\frac{1}{2 \beta}}{(L + 4 \ell_B /N)}^{-1}.
\end{equation}
This fits very well the exact evaluation for $N>10$ for $L < \ell_B$,
see Fig~(1a), even for $\lambda_s \ne 1$. Again for larger $L$ there
is an exponential decay in interactions.

\section{Interactions within an electrolyte}
To evaluate our expressions within an electrolyte we work in a
subspace corresponding to modes from \(-n_m:n_m\) of dimension
\(2n_m+1\). Using Matlab/Fortran labeling of the modes from 1 to
\(2n_m+1\), the mode \(m=0\) has the position \(i=n_m+1\) in the
matrix, and the mode \(m\) is at position \(i_m = n_m+m+1\). We can
then evaluate all the expressions using matrix algebra. $b$ is
evaluated using the matrix exponential and the free energies are
evaluated by grouping the combinatorial factors into right and left
vectors.

We now consider the case of a low ion concentration, $\bar
\lambda=0.02$ and take the chemical potential for charging the plates
as $\lambda_S=2$.  In figure~(\ref{fig:1} (a)) we plot the logarithm
of the absolute value of the disjoining pressure as a function of the
separation between the plates. We plot information on the charge state
of the surface: The constraint of strict neutrality has strong
influence on this behaviour.  For separations larger than $\sim 5
\ell_B$ the average charge of each plate (red dashed) is constant;
however for small separations the average surface charge goes to
zero. This we interpret as being due to the constraint of
electro-neutrality imposed by the chosen boundary conditions: The
system prefers to cancel the surface charge, rather than pulling in
counter-ions at small separations to ensure global electro-neutrality.

The fluctuations of the surface charge behave in a very different
manner from the average: The mean squared charge on the surface (blue
dotted) takes on a constant value for large separations, but {\bf
  increases strongly}\/ at smaller separations. Therefore, even though
on average the plates are neutral they can at any moment be strongly
charged. This is possible because there is a strong anti-correlation
between the two plates demonstrated in the evolution of the green
dashed-dotted curve -- one surface obviously becomes positive and the
other negative. Thus even though the average charge is driven to zero
there is a strong monopole fluctuation which we should interpret as
the Kirkwood-Shumaker effect \cite{kirkwood1, kirkwood2}.  The charge
cross-correlation can be seen to decrease to zero at large separations
due to the screening of electrostatic interaction by the electrolyte
and the short range attraction in this configuration can be
interpreted as being due to strong thermal monopole fluctuations.

Let us now consider a weakly coupled system with higher charge density
$\bar \lambda=10^3$. Rather rich behaviour is found as a function of
the surface properties. For $\lambda_S=2, N=5$ we have two changes of
sign of the disjoining pressure, figure~(\ref{fig:1} (a)),
\footnote{We plot pressures on logarithmic scales which gives
  characteristic singularities at sign changes.} We find repulsion at
both small and large separations, but a window of attraction at
intermediate distances. There is also a somewhat different behaviour
in the evolution of the charge state when compared to
figure~(\ref{fig:1} (b)) -- while the average charge is driven to zero
as before, we see that the blue fluctuations also decrease at small
separations so that monopole fluctuations are not strong enough to
produce short distance attraction.

If we change the charge state of the surface so that $N=1$ we find a
much simpler purely attractive interaction between plates. Again the
charge and fluctuations of the surface decrease at the smallest
separations, figure~(\ref{fig:1} (c)). It is interesting to note that
the amplitude of charge fluctuations remains high at all distances
even if the correlation between the plates is only important at the
smallest separations (when the sum of the blue and green curves is
driven to zero by neutrality).

Finally we force the surface to charge more strongly by increasing the
value of $\lambda_S=20$, figure~(\ref{fig:1} (d)). In this case the
average charge of the plates overwhelms the effects of
fluctuations. Even for this case the plates eventually discharge (on
average) at the smallest separation -- but the charge state again
fluctuates strongly.

We see from all of the figures in this section that the
electro-neutrality constraint has a strong influence on the charge
state, and charge correlations at small separations. The average
charge is always driven to zero leaving strong monopolar
fluctuations. In the next section we study a modified model in which
the electro-neutrality condition between the plates is relaxed so that
the surface can remain charged up to contact to see how this modifies
the interaction.

\begin{figure*}[t!]
  \begin{minipage}[b]{0.495\textwidth}\begin{center}
      \includegraphics[width=\textwidth]{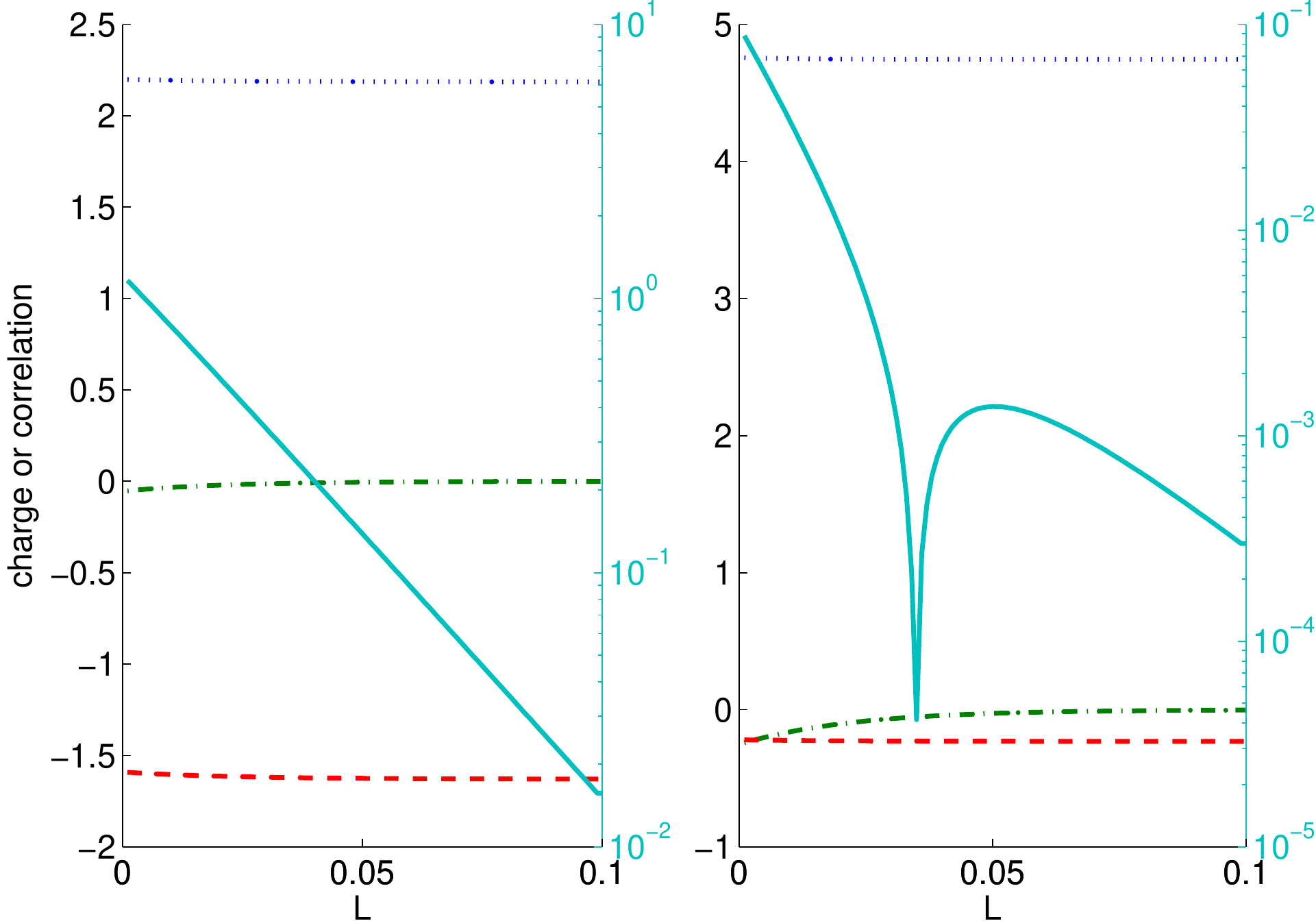}\\
      (a)~~~~~~~~~~~~~~~~~(b)
    \end{center}\end{minipage}\hskip0.1cm~~
  \begin{minipage}[b]{0.48\textwidth}\begin{center}
      \includegraphics[width=\textwidth]{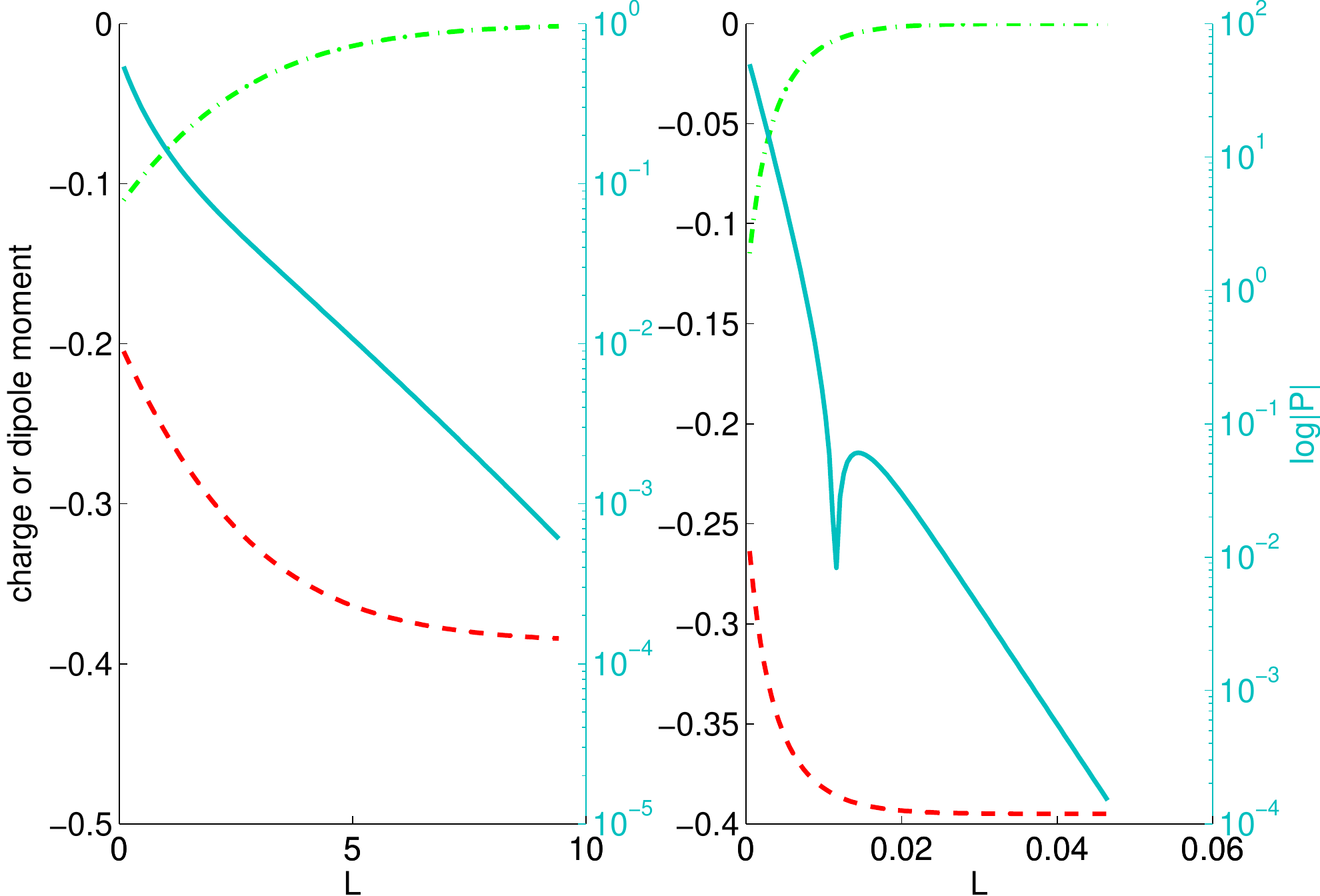}
      (c)~~~~~~~~~~~~~~~~~(d)
    \end{center}\end{minipage}
  \caption{Same color and curve coding as in
    figure~\ref{fig:1}. Plates within electrolyte. ({\bf a}):
    $\lambda_S=2$, $\bar \lambda=10^{3}$, $N=5$. The disjoining
    pressure is uniformly repulsive, but the charge state of the
    surface is stable as a function of the separation.  ({\bf b}): The
    small value of $\lambda_S=1.05$ only weakly charges the surface,
    but fluctuations remain very large. This leads to short range
    attraction between the surfaces. $N=10$. ({\bf c}): Disjoining
    pressure found to be attractive for all
    separations between charged dielectric "particles". The average charge (red dashed) and the dipole moment
    (green dashed-dotted) evolve as a function of the separation
    between the particles.  $\bar \lambda=10^{-1}$, $\lambda_S =2$, $
    N=5$.  ({\bf d}): Short range attractive, and long ranged
    repulsive interactions. The average charge (red dashed) evolves as
    a function of the separation as does the dipole moment of each
    particle (green dashed-dotted).  $\bar \lambda=10^3$, $\lambda_S
    =1.1$, $N=5$. }
  \label{fig:2}
\end{figure*}

\section{Two charged surfaces in an infinite electrolyte bath}

In this section we consider that the charges are attached to a
dielectric medium of zero thickness, with electrolyte on both sides of
the charged sources.  In this geometry we no longer force the average
charge of each plate to zero at contact; we can expect this to
increase the repulsion between the surfaces as a consequence.

The partition function, eq.~(\ref{partfun}), is now modified to
\begin{eqnarray}
  \Xi(L) = \lim_{D \rightarrow \infty} \int\; d\phi_0\; d\phi_L \; G_{\bar\lambda}(0, \phi_0; D) e^{-\beta f(0)} \nonumber\\
  G_{\bar\lambda}(\phi_0,
  \phi_L; L) e^{-\beta f(L)} G_{\bar\lambda}(\phi_L, 0; D),
  \label{partfun1}
\end{eqnarray}
where the electrolyte extends from $\lim_{D \rightarrow \infty} [-D,
D]$ and we assumed that the potential at the boundaries of this
interval vanishes.

In the low density regime figure (\ref{fig:2} (a)) shows that while in
this modified geometry the average charge on the plate does indeed
fall slightly at small separations, it remains finite at contact.  A
similar curve calculated for larger concentration of the electrolyte
shows that the charge state (both average and fluctuations) is only a
weak function of the separation. However it is interesting to notice
that the average charge is regulated by the chemical potentials of the
free and bound charges. While the average charge on the surface is far
from saturation, it resembles most clearly the standard boundary
condition of constant surface charge, independent of separation
between plates, for this set of parameters.

Finally we show that attraction is still possible, figure~(\ref{fig:2}
(b)) even in this geometry, if the chemical potential of the surface
is close to unity. In this case the average charge on the surface is
small and again fluctuations can be an important component of the
effective interaction. Note that the charge on each plate is strongly
fluctuating, but the plates remain weakly correlated for all
separations.

We next move on to consider a more sophisticated as well as a more
realistic model specifically for protein interaction, which
corresponds to a pair of interacting, dielectric particles with
ionizable surface groups, that is embedded in an infinite electrolyte.

\section{Two dielectric "particles" in an infinite electrolyte bath}

In this section we consider an idealization of a protein with
ionizable amino acids on its surface represented in a one dimensional
model. Each protein excludes the electrolyte and its interior core
behaves as a simple dielectric "particle" {with a dielectric constant
  different from its "bulk" value,} allowing in principle for the
inclusion of the polarization effects.

For simplicity we will take the dimension of this dielectric regions
as constant and equal to the Bjerrum length. {In general it depends on
  their actual thickness as well as on the dielectric discontinuity
  through the definition of the Bjerrum length.} The two surfaces of
each dielectric region are then described by our charge-regulated
model. We thus study a model of two pairs of charge-regulated
surfaces, each of them with a dielectric core, which interact through
an electrolyte solution. Apart from the obvious limitations of a one
dimensional model, this is as close as we can get to the realistic
description of the interactions between two proteins in an electrolyte
solution.

The partition function in this case is analogous to
eq.~(\ref{partfun1}) but we need to substitute
\begin{eqnarray}
  \label{field}
  &&\int\!\!d\phi_0 e^{-\beta f(0)} \longrightarrow \nonumber\\
  &&\int\!\!\!\!\int\!\!d\phi_{-h}d\phi_{h}\; e^{-\beta f(-h)} G_{0}(\phi_{-h}, \phi_{h}; 2h) e^{-\beta f(h)}
\end{eqnarray}
and
\begin{eqnarray}
  &&\int\!\!d\phi_L e^{-\beta f(L)} \longrightarrow \nonumber\\
  &&\int\!\!\!\!\int\!\!d\phi_{L-h}d\phi_{L+h}e^{-\beta f(L-h)}
  G_{0}(\phi_{L-h}, \phi_{L+h}; 2h) e^{-\beta f(L+h)}\nonumber\\
  .
\end{eqnarray}
where 
$G_{0}(\phi, \phi'; t)$ stands for the dielectric region without salt,
i.e., $\bar\lambda = 0$, representing the "particles" of which one
spans the interval $[-h, h]$ and the other one $[L-h, L+h]$. In our
numerical studies $2h$ equals one Bjerrum length.

It is clear that a number of interesting quantities can evolve as a
function of the separation of particles. As we have seen above, the
total charge of the particle will evolve, but it is also clear that
this will happen in an un-even manner for the charge facing the other
particle or facing the bulk electrolyte.  Thus there will also be a
dipole moment for each particle, which will naturally be
anti-correlated between the two particles and will increase in
amplitude at small separations. {This parallels very closely the
  original Kirkwood-Shumaker analysis \cite{kirkwood1, kirkwood2} that
  also considers separately monopolar fluctuations of the protein
  charge, as well as the associated fluctuations of the protein
  dipolar moment of the surface charge distribution.}

We firstly consider the interaction between two "particles" at low
electrolyte concentration {and therefore small screening},
figure~(\ref{fig:2} (c)). In this {region of the parameter space} we
find attraction at all separations between the particles, different from the previous case (figure~\ref{fig:2} (a)), where the disjoining pressure was uniformly repulsive.  The total charge and the dipole moment of the "particles"
show pronounced evolution as a function of separation, {with both
  in general decreasing in absolute value with the separation}.  The
attraction is generated predominantly by the
enhanced charge and dipolar moment correlations, originating from both "particles", while the 
average dipole moment of each "particle" is antisymmetric and contributes an additional repulsion.
  

The case of a denser electrolyte with small surface chemical potential
is presented in figure~(\ref{fig:2} (d)). As is already to be expected
from the simpler infinitely thin plate models, we generate
interactions with short ranged attraction and longer ranged
repulsion. Again we give the evolution of the charge and dipole states
of the particles both showing pronounced dependence on the separation.


\section{Conclusions}

We have demonstrated a rich variety of behaviours in a series of
models incorporating ionizable charge-regulated surfaces, i.e.,
surfaces that respond to the local electrostatic potential with a
variable effective charge, solved exacty in one dimension.

In the first model, of a pair of chargeable metallic plates, an
important part of the physics comes from the possibility of driving
the surface charge to zero at small separation. When this happens,
large, correlated fluctuations occur which can lead to attractive
interactions of a Kirkwood-Shumaker type.

A second model of two neutral plates in an infinite electrolyte bath
considerably weakens the effect of global charge neutrality. But with
chemical potentials close to unity, implying small charge dissociation
energy penalty, fluctuation effects can still be very strong at the
surfaces, affecting their interaction.

A third model is inspired directly by the work of Kirkwood-Shumaker on
the effect of charge fluctuations and their role in the interactions
between proteins, here idealised as dielectric "particles" of
finite size with dissociable surface charge groups. This model takes into account the charge regulation of
dissociable groups (amino acids) on the surface as well as the
fluctuations of the mono polar and dipolar components of the
fluctuating charge distribution, being thus closest to the original
  Kirkwood-Shumaker proposition \cite{kirkwood1, kirkwood2}. 

We expect that at least some of the properties of these models will
transfer also to the more realistic 3D models of the interaction
between globular proteins with dissociable surface charge groups. {Our
  analysis suggests that there might be features of electrostatic
  interactions between macro ions bearing dissociable charged groups
  that have heretofore not been specifically considered in the models
  of protein-protein interaction \cite{biochargereg}.}

\section{Acknowledgment}

A.C.M. is financed by the ANR grant FSCF. R.P.  thanks the hospitality
of ESPCI during his stay in Paris as a visiting professor and
acknowledges ARRS grant P1-0055.

\bibliographystyle{eplbib} \bibliography{rudip}

\end{document}